\documentclass[10pt,twocolumn,letterpaper]{article}

\usepackage{iccv}              

\usepackage{algorithm}
\usepackage{algpseudocode}

\definecolor{iccvblue}{rgb}{0.21,0.49,0.74}
\usepackage[pagebackref,breaklinks,colorlinks,allcolors=iccvblue]{hyperref}

\def\paperID{***} 
\def\confName{ICCV}
\def\confYear{2025}

\title{MeshSplat: Generalizable Sparse-View Surface Reconstruction via Gaussian Splatting}

\author{%
  Hanzhi Chang$^{1}$\footnotemark[1], \quad Ruijie Zhu$^{1,2}$\footnotemark[1], \quad Wenjie Chang$^{1}$, \quad Mulin Yu$^{2}$, \\
  Yanzhe Liang$^{1}$,  \quad Jiahao Lu$^{1}$, \quad Zhuoyuan Li$^{1}$, \quad Tianzhu Zhang$^{1}$\footnotemark[2]\ \\\\
  $^1$University of Science and Technology of China\;\;
  $^2$Shanghai AI Laboratory\\
}

\begin{document}

\twocolumn[{%
\maketitle
\begin{figure}[H]
\hsize=\textwidth
\centering
\includegraphics[width=17.5cm]{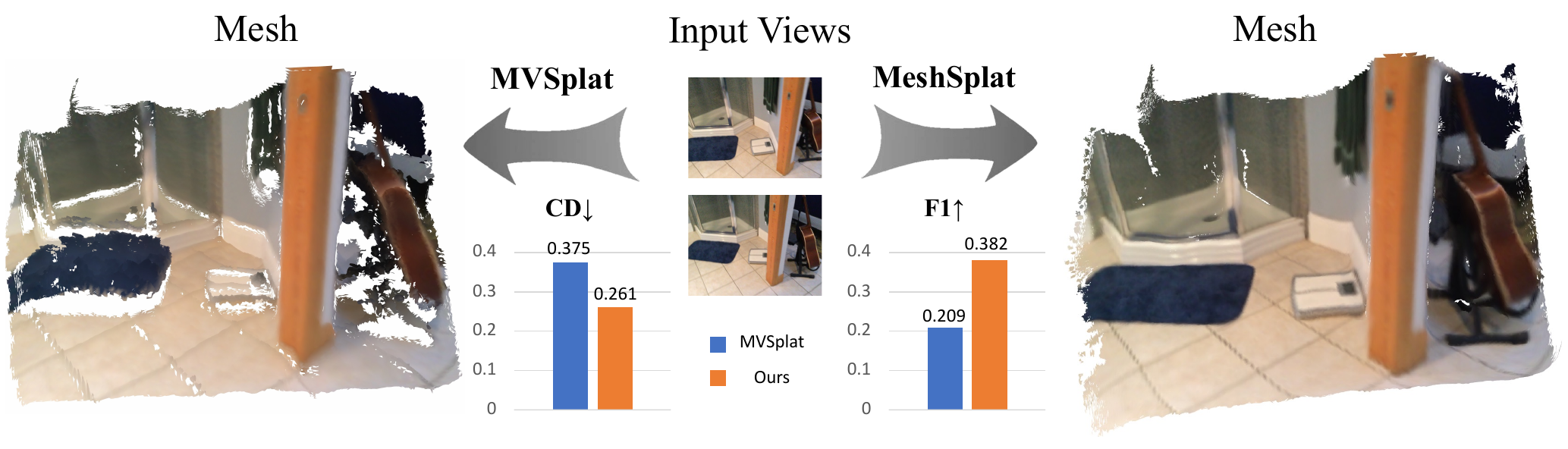}
\caption{Given sparse-view images as input, \textbf{MeshSplat} can directly predict the scene geometry and efficiently extract the scene mesh. Compared to MVSplat~\cite{chen2024mvsplat} and other state-of-the-art methods, Meshplat achieves more consistent and precise mesh extraction in \textbf{generalizable sparse-view surface reconstruction.}}
\end{figure}
}]

{
\renewcommand{\thefootnote}{\fnsymbol{footnote}}
\footnotetext[1]{Equal contributions.}
\footnotetext[2]{Corresponding author.}
}

\begin{abstract}
Surface reconstruction has been widely studied in computer vision and graphics. However, existing surface reconstruction works struggle to recover accurate scene geometry when the input views are extremely sparse. To address this issue, we propose MeshSplat, a generalizable sparse-view surface reconstruction framework via Gaussian Splatting. Our key idea is to leverage 2DGS as a bridge, which connects novel view synthesis to learned geometric priors and then transfers these priors to achieve surface reconstruction. Specifically, we incorporate a feed-forward network to predict per-view pixel-aligned 2DGS, which enables the network to synthesize novel view images and thus eliminates the need for direct 3D ground-truth supervision. To improve the accuracy of 2DGS position and orientation prediction, we propose a Weighted Chamfer Distance Loss to regularize the depth maps, especially in overlapping areas of input views, and also a normal prediction network to align the orientation of 2DGS with normal vectors predicted by a monocular normal estimator. Extensive experiments validate the effectiveness of our proposed improvement, demonstrating that our method achieves state-of-the-art performance in generalizable sparse-view mesh reconstruction tasks. Project Page: \url{https://hanzhichang.github.io/meshsplat_web/}.
\end{abstract}    
\section{Introduction}
\label{sec:intro}

Surface reconstruction of 3D scenes is a fundamental task in 3D vision, with a wide range of applications in downstream tasks such as AR/VR and embodied AI. Recently, based on NeRF~\cite{mildenhall2021nerf} and 3DGS~\cite{kerbl20233d}, numerous per-scene optimized surface reconstruction methods~\cite{wang2021neus,huang20242d,chen2023neusg,yu2024gsdf,guedon2024sugar} have been proposed. However, these methods struggle to robustly reconstruct scenes with only sparse views as input. This is because sparse views can only provide limited multi-view geometric constraints, which are insufficient to perform high-quality per-scene geometry optimization. Therefore, it is necessary to construct a feed-forward pipeline to learn geometric priors from diverse scenarios for generalizable sparse-view surface reconstruction.

To address this issue, several meaningful attempts have been made, which are mainly NeuS-based methods~\cite{wang2021neus}. For instance, SparseNeuS~\cite{long2022sparseneus} constructs geometric volumes to estimate implicit SDF fields, which are then transformed to mesh. However, due to the inefficiency of implicit SDF representation and neural rendering, NeuS-based methods are limited to object-centric scenes and suffer from long rendering times.
Compared with NeuS, Gaussian splatting can achieve not only faster rendering speeds but also better rendering quality. However, the vanilla Gaussian splatting framework falls short in representing surface geometry~\cite{huang20242d}. Therefore, existing Gaussian splatting-based methods~\cite{charatan2024pixelsplat,chen2024mvsplat,wewer2024latentsplat} mainly focus on generalizable novel view synthesis, while methods applying Gaussian splatting to generalizable sparse-view surface reconstruction remain unexplored.

\begin{figure}
    \centering
    \includegraphics[width=1.0\linewidth]{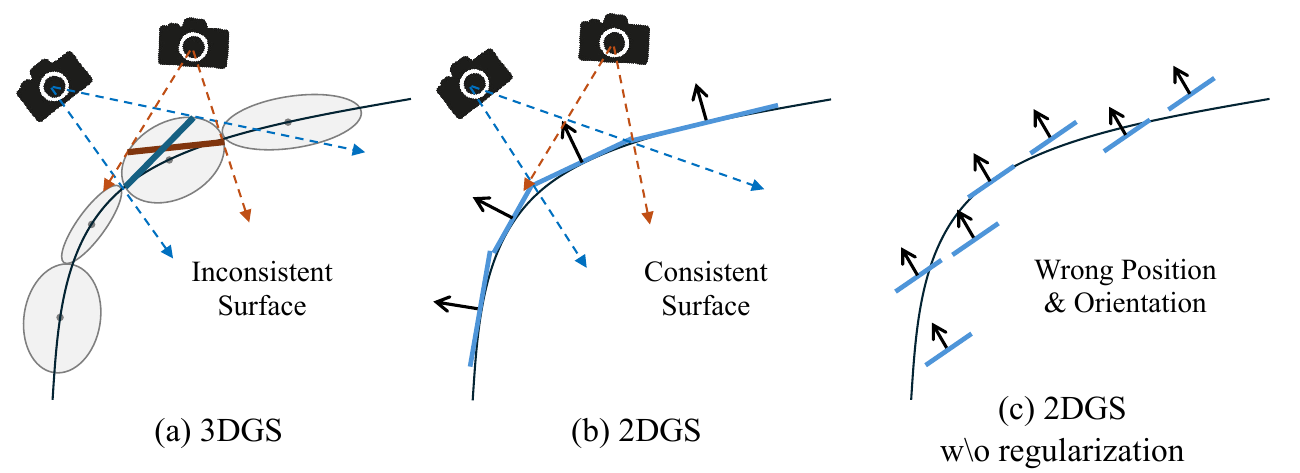}
    \caption{\textbf{Motivation.} (a) The ellipsoid shape of 3DGS leads to different intersection planes in different viewpoints, resulting in inconsistent surface. (b) 2DGS has consistent intersection planes in different viewpoints, which is more suitable for surface reconstruction. (c) When the positions and orientations of 2DGS are not regularized, there will be significant discrepancies between 2DGS and the contours of the surface, which hinders the reconstruction of scene surfaces.}
    \label{fig:2dgs}
\end{figure}

To realize high-quality surface reconstruction, the network requires not only realistic appearance rendering but also precise geometry recovery. More importantly, the former task usually has more available training data, while the latter requires expensive geometric labeling. This raises a question: \textbf{is it possible to learn generalizable geometric priors from novel view synthesis task?} Motivated by recent work 2DGS~\cite{huang20242d}, we find an effective way to address this issue: using 2DGS as a bridge between novel view synthesis (NVS) and surface reconstruction. Compared with 3DGS, 2DGS is naturally more suitable for representing thin surfaces and can be easily used to extract surface mesh beyond rendering novel views, as shown in Figure~\ref{fig:2dgs} (a) and (b).
To this end, we decide to construct a feed-forward framework to predict pixel-aligned 2DGS for generalizable sparse-view surface reconstruction.

However, we also find that integrating 2DGS into generalizable sparse-view 3D scene reconstruction is not a trivial task. Compared to 3DGS, 2DGS is more sensitive to \textbf{position} and \textbf{orientation} estimation in mesh reconstruction, as illustrated in Figure~\ref{fig:2dgs} (c).
For position, it can be iteratively optimized in a per-scene optimization setting. However, in an end-to-end framework, the Gaussian positions depend on the predicted depth maps. Due to the thin nature of 2DGS, mispredictions in the depth map directly lead to noticeable position shifts, unlike 3DGS.
For orientation, unlike 3DGS whose orientations can be more flexible, the orientation of 2DGS directly determines the surface normals of the scene. Errors in predicting Gaussian orientation will ultimately result in distorted scene surfaces.

To address these challenges, we propose \textbf{MeshSplat}, a generalizable sparse-view surface reconstruction framework based on 2DGS, which consists of the following key designs:
(1) Given sparse images as input, we build an \textbf{MVS-based feed-forward network to generate pixel-aligned 2DGS}, where the position and orientation of 2DGS are transformed according to the predicted depth and normal maps, and other attributes of 2DGS are decoded directly from a convolutional Gaussian head. With the predicted 2DGS, we can synthesize novel views for supervision and finally perform mesh extraction.
(2) To improve the accuracy of 2DGS position prediction, we design a \textbf{Weighted Chamfer Distance Loss} to align point clouds generated from predicted depths across different views. Since adjacent views have varying overlaps, we introduce a confidence map to identify overlapping regions and weight the Chamfer Distance Loss accordingly, assigning higher weights to these overlapping areas.
(3) To improve the accuracy of 2DGS orientation prediction, we design an \textbf{uncertainty-based normal prediction network} that predicts per-view normals and converts them to rotation quaternions for 2DGS. The per-view normals are supervised by an off-the-shelf monocular normal estimator, which incorporates monocular geometry priors to assist surface reconstruction.

Our main contributions can be summarized as follows:
\begin{itemize}
    \item We propose MeshSplat, the first method, to the best of our knowledge, that focuses on generalizable sparse-view surface reconstruction via Gaussian splatting. 
    \item By incorporating 2DGS instead of 3DGS as a bridge, MeshSplat can learn generalizable geometric priors from the novel view synthesis task in a self-supervised manner.
    \item To better integrate 2dgs into the feed-forward framework, we introduce a Weighted Chamfer Distance Loss to align Gaussian position across views and design an uncertainty-based normal prediction network that provides monocular supervision to Gaussian orientations.
    \item Extensive experiments show that our method achieves state-of-the-art performance in generalizable sparse-view surface reconstruction and cross-dataset generalization.
\end{itemize}
\section{Related Work}
\label{sec:related}

\subsection{3D Scene Representation}

Early 3D scene methods~\cite{jiang2020local,genova2020local,cheng2025bridge} employ point clouds, signed distance functions or occupancy fields to represent 3D scenes. NeRF~\cite{mildenhall2021nerf} utilizes neural networks to construct the 3D radiance field for scene representation and renders RGB images by volume rendering, which inspires many subsequent works in 3D vision~\cite{barron2021mip,muller2022instant,honglrm}. However, NeRF-based methods suffer from long-time inefficient rendering. 3DGS~\cite{kerbl20233d} models 3D scenes by explicit ellipsoidal Gaussian primitives and introduces a differentiable rendering method based on alpha-blending. It has become the core 3D scene representation for many tasks, such as novel view synthesis~\cite{ren2025octree,lu2024scaffold}, 3D assets generation~\cite{tangdreamgaussian} and dynamic scene reconstruction~\cite{huang2024sc,zhu2024motiongs}. Despite their success, original NeRF and 3DGS require per-scene optimization, lacking cross-scene generalization capabilities.

\subsection{Generalizable 3D Reconstruction}

Extensive works have proposed generalizable 3D reconstruction frameworks by training neural networks on large-scale datasets to learn cross-scene priors. Early approaches~\cite{chen2021mvsnerf,wang2021ibrnet,xu2024murf} primarily combined feed-forward neural networks with Neural Radiance Fields (NeRF). However, these methods often suffer from the inefficiencies inherent in NeRF training and rendering, which limit their performance. 
More recently, methods such as DUSt3R and its subsequent extensions~\cite{wang2024dust3r,duisterhof2025mast3r,wang2025vggt} leverage Transformer-based architectures trained on large-scale 3D-annotated datasets to predict 3D point maps from input 2D images. While effective in generating point-based 3D representations, these approaches cannot generalize to novel views and are not suitable for surface reconstruction tasks.
Methods like LVSM~\cite{jinlvsm} support novel view synthesis (NVS) tasks, but they are limited to generating RGB images and thus cannot be directly applied to surface reconstruction.
With the introduction of 3D Gaussian Splatting (3DGS), a new class of methods~\cite{charatan2024pixelsplat,chen2024mvsplat,wang2024freesplat,wang2025zpressor}, where MVSplat~\cite{chen2024mvsplat} represents a prominent approach, employ 3DGS as the 3D scene representation for generalizable 3D reconstruction.  They usually first predict depth maps from input views, and then use back-projected 3D points as centers to construct pixel-aligned 3D Gaussian representations. PM loss~\cite{shi2025revisiting} further regularizes the predicted 3DGS positions by introducing VGGT priors. However, despite their promising performance in NVS tasks, these generalizable frameworks remain inadequate for mesh extraction and surface reconstruction tasks.

\subsection{Surface Reconstruction}

Traditional surface reconstruction methods~\cite{kazhdan2013screened,furukawa2009accurate} primarily model 3D scene geometry through explicit feature matching. With the rise of neural networks, many works~\cite{wang2021neus,reiser2024binary,li2023neuralangelo,yariv2021volume} reconstruct scene surface by estimating neural occupancy fields or signed distance functions (SDF) by neural networks, and render novel views through surface rendering or volume rendering techniques. Recently, many works have extended Gaussian Splatting to mesh extraction. These methods can be categorized into two types: methods that directly use ellipsoidal 3DGS to extract meshes~\cite{guedon2024sugar,yu2024gsdf,yu2024gaussian,zhu2025objectgs}, and methods that use flattened 3DGS or 2DGS~\cite{huang20242d,chen2023neusg,chen2024pgsr}. However, these methods fail to generate high-quality meshes under sparse viewpoint inputs and lack cross-scene generalization.
SparseNeuS~\cite{long2022sparseneus} and subsequent works~\cite{peng2023gens,xu2023c2f2neus,chen2024neuralrecon}, inspired by generalizable novel view synthesis approaches, achieve generalizable surface reconstruction based on NeuS~\cite{wang2021neus}. These methods extract feature maps from input images to construct a 3D geometry volume, which is used to obtain neural SDF fields and output the mesh of the scene. However, these methods exhibit poor efficiency in mesh extraction and are limited to object-centric scenes. 2DGS~\cite{huang20242d} outperforms NeuS-based methods in mesh extraction tasks, but how to generalize 2DGS to different scenes with sparse inputs still remains unexplored. Therefore, we propose MeshSplat, an end-to-end generalizable sparse-view surface reconstruction framework based on 2DGS.

\section{Methods}

\begin{figure*}
  \centering
  \includegraphics[width=0.95\linewidth]{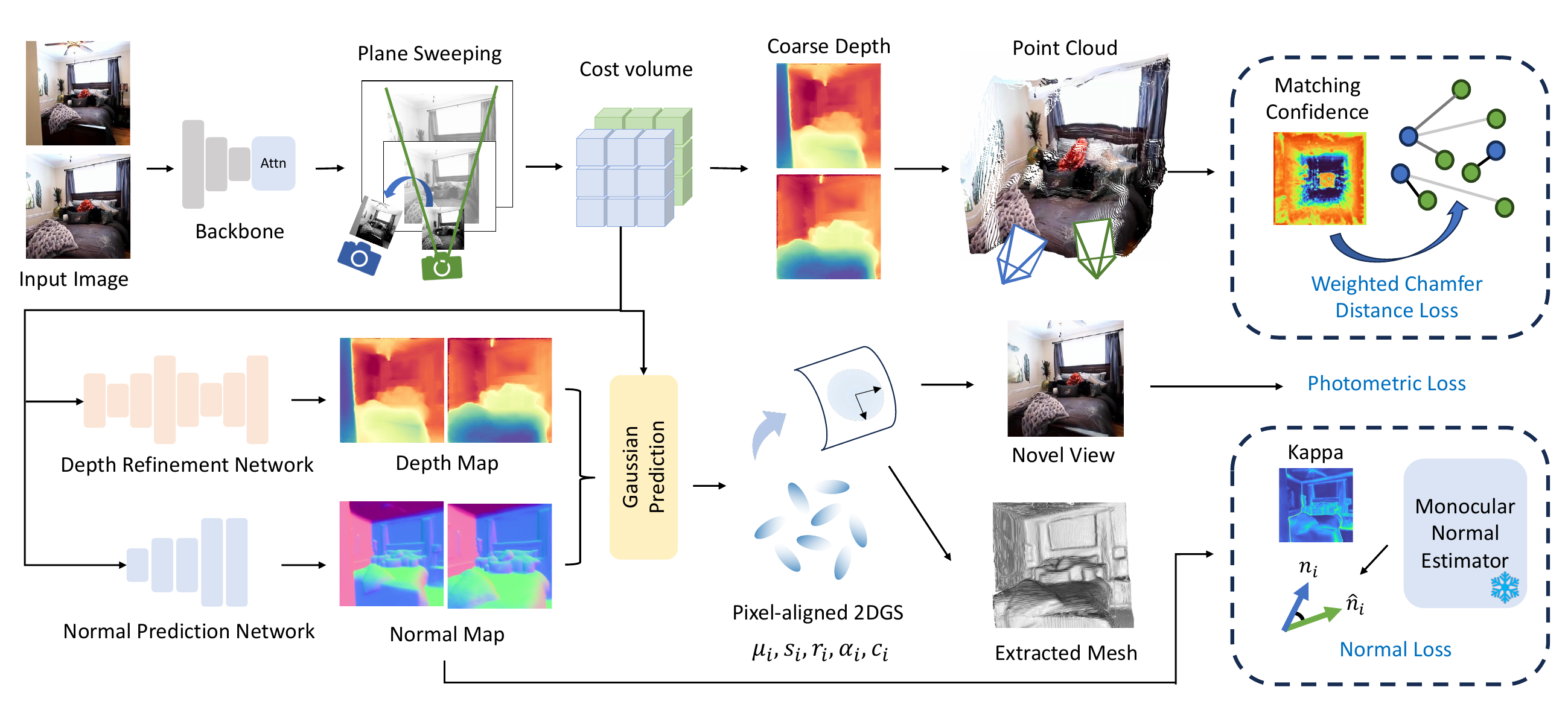}
  \caption{\textbf{Overall Architecture.} Taken a pair of images as input, MeshSplat begins with a multi-view backbone to extract per-view feature maps. Then we construct per-view cost volumes via the plane-sweeping to generate coarse depth maps, which can be projected to 3D point clouds and be constrained by our proposed Weighted Chamfer Distance Loss. We further feed cost volumes into our gaussian prediction network, together with a depth refinement network and a normal prediction network, to obtain pixel-aligned 2DGS. Finally, we use these 2DGS to render novel view for supervision and reconstruct the scene mesh.}
  \label{fig:Methods}
\end{figure*}

\subsection{Overall Framework}
\label{sec:methods_framework}

The framework of MeshSplat is illustrated in Figure~\ref{fig:Methods}. Given two images $\{I_i\}_{i=1}^2$ and corresponding projection matrices$\{\Pi_i\}_{i=1}^2,\ \Pi_i=K_i[R_i|t_i]$, where $K_i$ denotes for camera intrinsics, $R_i$ denotes for camera rotation matrices and $t_i$ denotes for translation matrices, MeshSplat begins with an encoder composed of a CNN and Multi-View Transformer to extract feature maps. From these multi-view feature maps, per-view cost volumes are constructed via plane sweeping. To enhance the multi-view consistency, we propose a Weighted Chamfer Distance Loss to constrain the cost volumes. Subsequently, we use a Gaussian Prediction Network, which includes a normal prediction network to predict Gaussian orientations, and a depth refinement network to predict Gaussian positions. In this way, we can obtain the pixel-aligned 2DGS to represent the scene. This process can be formulated as:
\begin{equation}
  \{I_i,\Pi_i\}_{i=1}^2 \rightarrow \{\mu_j, s_j, r_j, \alpha_j, c_j\}_{j=1}^{2\times H\times W}
  \label{eq:formulation}
\end{equation}
Finally, these predicted 2DGS can render novel views and reconstruct the scene mesh.

\subsection{Cost Volume Construction}
\label{sec:methods_costvol}

Following MVSplat~\cite{chen2024mvsplat}, we first use a CNN to extract a feature map from each input view. Then, we employ a Transformer that incorporates multi-view cross-attention to enable information exchanges across different views. 
As a result, we obtain the feature map $\{F_i\}_{i=1}^2$ for each view with a size of $\frac{H}4\times \frac{W}{4}\times C$.

To intuitively obtain the matching correspondences between the two input images, we construct per-view cost volumes via plane sweeping. Specifically, for input view $i$, we first divide the depth ranges into several depth bins $\{d_k\}_{k=1}^D$. The feature map $F_j$ from the other view $j$ is then warped to view $i$ using the projection matrices $P_i$ and $P_j$ along with the current depth candidate $d_k$, denoted as $F_{j \rightarrow i}^{d_k}$. Next, we compute the dot product between the two feature maps and concatenate the results from all depth candidates. After two convolutional layers $\phi_{CV}$, we finally obtain the cost volume $V_i$ with a size of $\frac H 4 \times \frac W 4\times D$:
\begin{equation}
  V_i=\phi_\mathrm {CV}(\{V_i^{d_k}\}_{k=1}^D),\ \ \ V_i^{d_k}=\frac{F_i \cdot F_{j\rightarrow i}^{d_k}}{\sqrt C}
  \label{eq:costvolume}
\end{equation}

Since the correlation calculation directly reflects the matching confidence between the input images, we can directly apply Softmax operation to compute the probabilities of the depth values to be each depth candidates. The coarse depth maps can be obtained by weighted summation:
\begin{equation}
  W_i = \mathrm {Softmax}_D(V_i),\ \ \ D^\mathrm {coarse}_i=\sum_kW_i^{k}d_k
  \label{eq:depth}
\end{equation}

\subsection{Gaussian Prediction Network}
\label{sec:methods_gspredict}

Next, we discuss how to derive the attributes of pixel-aligned 2DGS from cost volumes. Following MVSplat~\cite{chen2024mvsplat}, we first use U-Net with attention layers to refine the cost volume, taking feature maps as additional inputs. 
Using the refined cost volume, we can compute the final depth maps using equation~\eqref{eq:depth}, and then perform back-projection to obtain the 3D positions of the pixel-aligned 2DGS. 

In addition, 2DGS requires more precise orientation prediction than the ellipsoid-like 3DGS. Based on this, we introduce an additional normal prediction network to predict the normal maps of the input images, which are further converted to the 2DGS rotation quaternions. We use a lightweight CNN $\phi_\mathrm {rot}$ to predict the orientations of the pixel-aligned 2DGS. The network takes the refined cost volumes, feature maps, and original RGB images at three different scales as input, and processes them through convolutional layers with residual connections. Finally the rotation quaternions $q$ for all three scales are obtained, which are further converted into normal vectors $n$ for each scale, along with kappa $\kappa$ which reflect the uncertainty of predicted normal vectors following \cite{bae2021estimating}:
\begin{equation}
  \{q,\kappa\}=\phi_\mathrm {rot}(V_i||F_i||I_i),\ \ \ n=R(q)\cdot[0,0,1]^T
  \label{eq:normalpred}
\end{equation}
where $R(q)$ represents the rotation matrix computed from the rotation quaternion $q$. The resulting normal vectors $n$ and kappa $\kappa$ will be used in subsequent loss calculations.

For the remaining attributes of 2DGS, following \cite{chen2024mvsplat}, we predict them through a two-layer convolutional gaussian head, taking the refined cost volumes, feature maps, and original RGB images as input. 

\subsection{Weighted Chamfer Distance Loss}
\label{wcdloss}

In the ideal case, the Gaussian positions predicted from adjacent views should have significant overlaps. Considering this property, one approach is to use the Chamfer Distance (CD) Loss as regularization on these 3D points. Chamfer Distance measures the overlaps between two point clouds, which can be expressed as:
\begin{equation}
  \mathcal L_\mathrm {CD}=\frac{1}{2}(\frac{1}{N_1}\sum_{i=1}^{N_1}\min_{j}||p_1^i-p_2^j||+\frac{1}{N_2}\sum_{i=1}^{N_2}\min_j||p_2^i-p_1^j||)
  \label{eq:cdloss}
\end{equation}
where $\{p_k^i\}_{i=1}^{N_k}$ are the predicted point clouds calculated by the projection matrices $\Pi_k$ and coarse depth maps $D^{\mathrm{coarse}}_k$ of the $k$-th view. 
The original CD Loss assigns equal weights to all points. However, due to occlusion and view differences, those pixels which do not have their corresponding pixels will have far chamfer distances. Applying CD loss uniformly to all points may result in incorrect and unreasonable constraints. 

To address this issue, we design confidence maps to measure the matching confidence of each pixel and use them to weight CD loss, resulting in our proposed Weighted Chamfer Distance Loss (WCD Loss). We argue that confidence maps can be derived from the cost volumes, since they have already captured feature similarities between pixels. The confidence map $M_i$ for view $i$ is defined by taking the maximum value along depth dimension in the cost volume:
\begin{equation}
  M_i = \max_{d_k} \mathrm {Softmax}_D(V_i),\ \ \ i=1,2
  \label{eq:uncertainty_map}
\end{equation}
With the confidence maps, we define the WCD Loss as:
\begin{equation}
\begin{split}
  \mathcal L_\mathrm {WCD}=\frac{1}{2}(\frac{1}{N_1}\sum_{i=1}^{N_1}M_1(i)\min_{j}||p_1^i-p_2^j||\\
  +\frac{1}{N_2}\sum_{i=1}^{N_2}M_2(i)\min_j||p_2^i-p_1^j||)
  \label{eq:wcdloss}
\end{split}
\end{equation}
where $M_{1/2}(i)$ denotes the confidence map value for pixel $i$ in the first or second view.

Our proposed WCD loss ensures that the regions with higher matching confidence overlap as much as possible. This regularizes the generation of cost volumes, and leads to a more accurate prediction of 2DGS positions.

\subsection{Uncertainty-Guided NLL Normal Loss}
\label{ugnll_loss}

Due to the lack of regularization for normals, the predicted orientations of 2DGS still suffer from degradation, preventing the 2DGS from aligning precisely with the actual surfaces. To address this issue, we use the uncertainty-guided normal negative log-likelihood (NLL) loss proposed by \cite{bae2021estimating} to supervise predicted normal maps at each scale, which is defined as:

\begin{equation}
\begin{split}
  \mathcal L_\mathrm {AngMF}(n_i,\hat n_i,\kappa_i)=-\log(\kappa_i^2+1)+\log(1+\\\exp(-\kappa_i\pi))+\kappa_i\cos^{-1}n_i^T\hat n_{i}
\end{split}
\label{eq:angmf_loss}
\end{equation}
where $n_i$, $\kappa_i$, $\hat n_i$ stands for the predicted normal vector, kappa and pseudo ground-truth normal vector of pixel $i$. Following \cite{bae2021estimating}, we sample pixels based on their $\kappa_i$, forming the sampled sets $P_\mathrm {sample}$, and only apply NLL loss on $P_\mathrm {sample}$. We use an off-the-shelf monocular normal estimation model Omnidata~\cite{eftekhar2021omnidata} to provide pseudo ground-truth normal maps as supervision. Finally, the normal loss $\mathcal L_\mathrm {normal}$ is computed as the average of NLL losses in sampled points. Details can be found in the Appendix.

\subsection{Training and Inference}
\label{training}
Finally, our overall training loss is defined as:
\begin{equation}
  \mathcal L=w_1\mathcal L_\mathrm {pho}+w_2\mathcal L_\mathrm {WCD}+w_3\mathcal L_{\mathrm {normal}}
  \label{eq:total_loss}
\end{equation}
$\mathcal L_\mathrm {pho}$ consists of MSE loss and LPIPS loss calculated from ground truth RGB image $I$ and rendered image $\hat I$:
\begin{equation}
  \mathcal L_\mathrm {pho}=w_{11}\ \mathrm{MSE}(I,\hat I)+w_{12}\ \mathrm{LPIPS}(I,\hat I)
  \label{eq:photo_loss}
\end{equation}

During inference, 2DGS is first reconstructed through our network. Then we follow 2DGS original paper~\cite{huang20242d} to extract the scene mesh.

\begin{table*}[t]
\centering
\caption{\textbf{Quantitative Comparisons.} * denotes that we use the dense reconstruction results of COLMAP with dense view inputs as ground-truth point clouds of Re10K. ** denotes for the 300k-training-step version according to its original setting. }
\begin{tabular}{lccccccccc}
\toprule
           & \multicolumn{4}{c}{Re10K*}             & & \multicolumn{4}{c}{Scannet}              \\ \cmidrule{2-5}  \cmidrule{7-10}
           & CD↓         & Precision↑  & Recall↑     & F1↑         & & CD↓         & Precision↑  & Recall↑     & F1↑      \\ \midrule
SparseNeuS~\cite{long2022sparseneus} & 6.0473           & 0.0012           & 0.0097           & 0.0020          & & 7.1860           & 0.0056           & 0.1964           & 0.0107           \\
MVSNeRF~\cite{chen2021mvsnerf}    & 0.6139           & 0.1390           & 0.1548           & 0.1407          & & 0.5761           & 0.1320           &  0.2053          & 0.1514           \\
pixelSplat~\cite{charatan2024pixelsplat} & 1.4423           & 0.1067           & 0.0903           & 0.0944          & & 0.3285 & 0.2599 & 0.3597 & 0.2948            \\
MVSplat**~\cite{chen2024mvsplat}   & 0.4038   & 0.3986          & 0.2633       & 0.3139  & & - & - & - & -           \\
MVSplat~\cite{chen2024mvsplat}    & 0.4015   & 0.3949          & 0.2607       & 0.3100  & & 0.3748 & 0.1992 & 0.2282 & 0.2095           \\
MeshSplat (Ours) & \textbf{0.3566}   & \textbf{0.5289}          & \textbf{0.2953}       & \textbf{0.3758}  & & \textbf{0.2606} & \textbf{0.3901} & \textbf{0.3849} & \textbf{0.3824}           \\ \bottomrule
\end{tabular}

\label{tab:re10k}
\end{table*}

\begin{table*}[t]
\centering
\caption{\textbf{Quantitative Comparisons in Zero-Shot Transfer Experiments on Scannet and Replica Datasets.} We use the models trained on Re10K only to perform cross-dataset generalization experiments. MeshSplat still shows promising results. }
\begin{tabular}{lccccccccc}
\toprule
           & \multicolumn{4}{c}{Re10K $\rightarrow$ Scannet}             & & \multicolumn{4}{c}{Re10K $\rightarrow$ Replica}              \\ \cmidrule{2-5} \cmidrule{7-10} 
           & CD↓         & Precision↑  & Recall↑     & F1↑        & & CD↓         & Precision↑  & Recall↑     & F1↑         \\ \midrule
SparseNeuS~\cite{long2022sparseneus} & 8.3055           & 0.0003           & 0.0166           & 0.0006          & & 12.72           & 0.0002           & 0.0099 & 0.0003           \\
MVSplat~\cite{chen2024mvsplat}    & 0.4506 & 0.1333 & 0.1591  & 0.1418 & & 1.089 & 0.0528 & 0.0695 & 0.0564  \\
MeshSplat (Ours)  & \textbf{0.3148} & \textbf{0.2868} & \textbf{0.3176} & \textbf{0.2956} & & \textbf{0.921} & \textbf{0.0749} & \textbf{0.0984} & \textbf{0.0809} \\ 
\bottomrule
\end{tabular}

\label{tab:zero_shot}
\end{table*}

\begin{figure*}[t!]
  \centering
  \includegraphics[width=\linewidth]{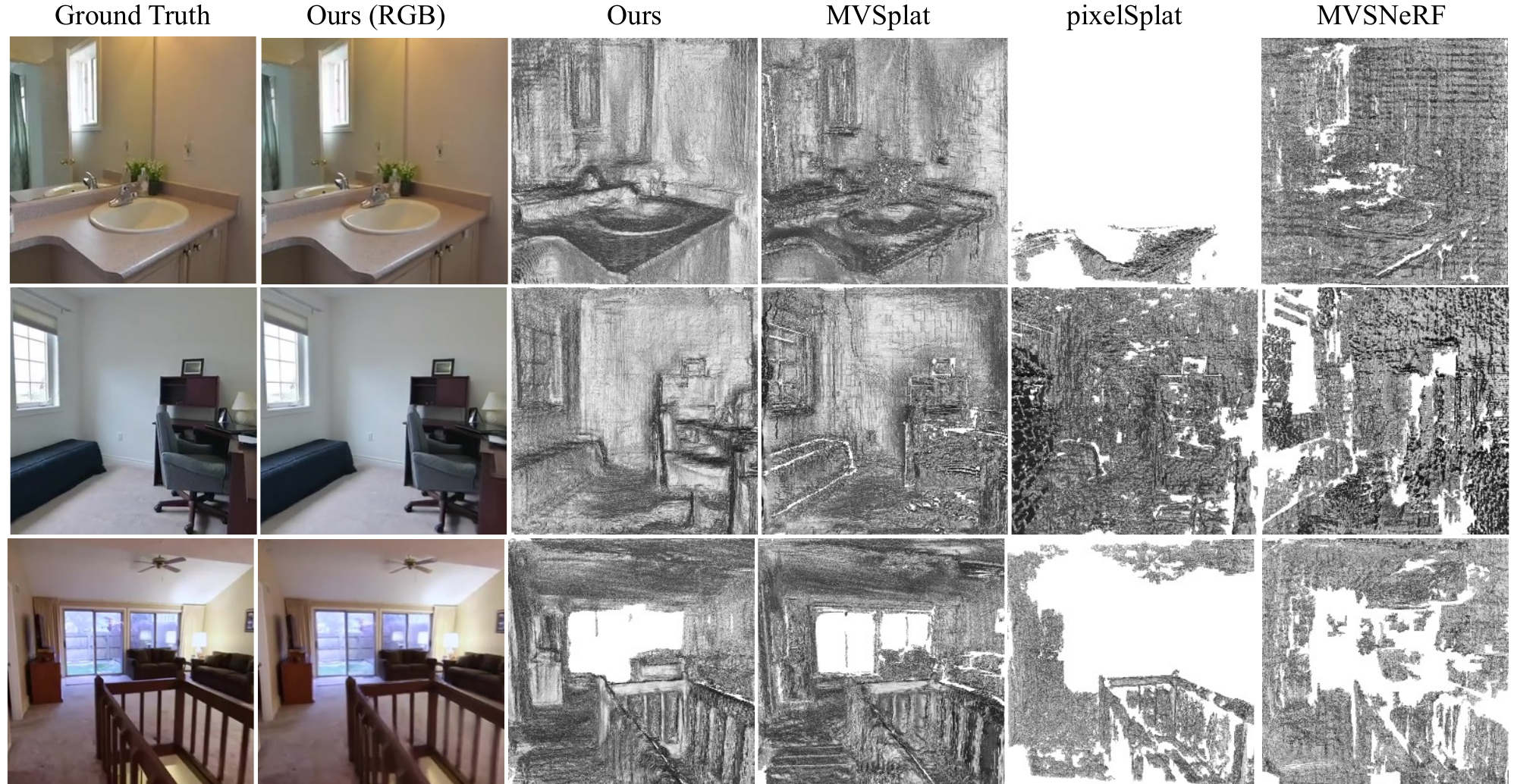}
  \caption{\textbf{Quanlitative Comparisons on Re10K Dataset.} While the baseline methods provide meshes with holes and uneven surfaces, MeshSplat successfully reconstruct the scene with smoother and more complete meshes. }
  \label{fig:main_re10k}
\vspace{1em}
  \centering
  \includegraphics[width=\linewidth]{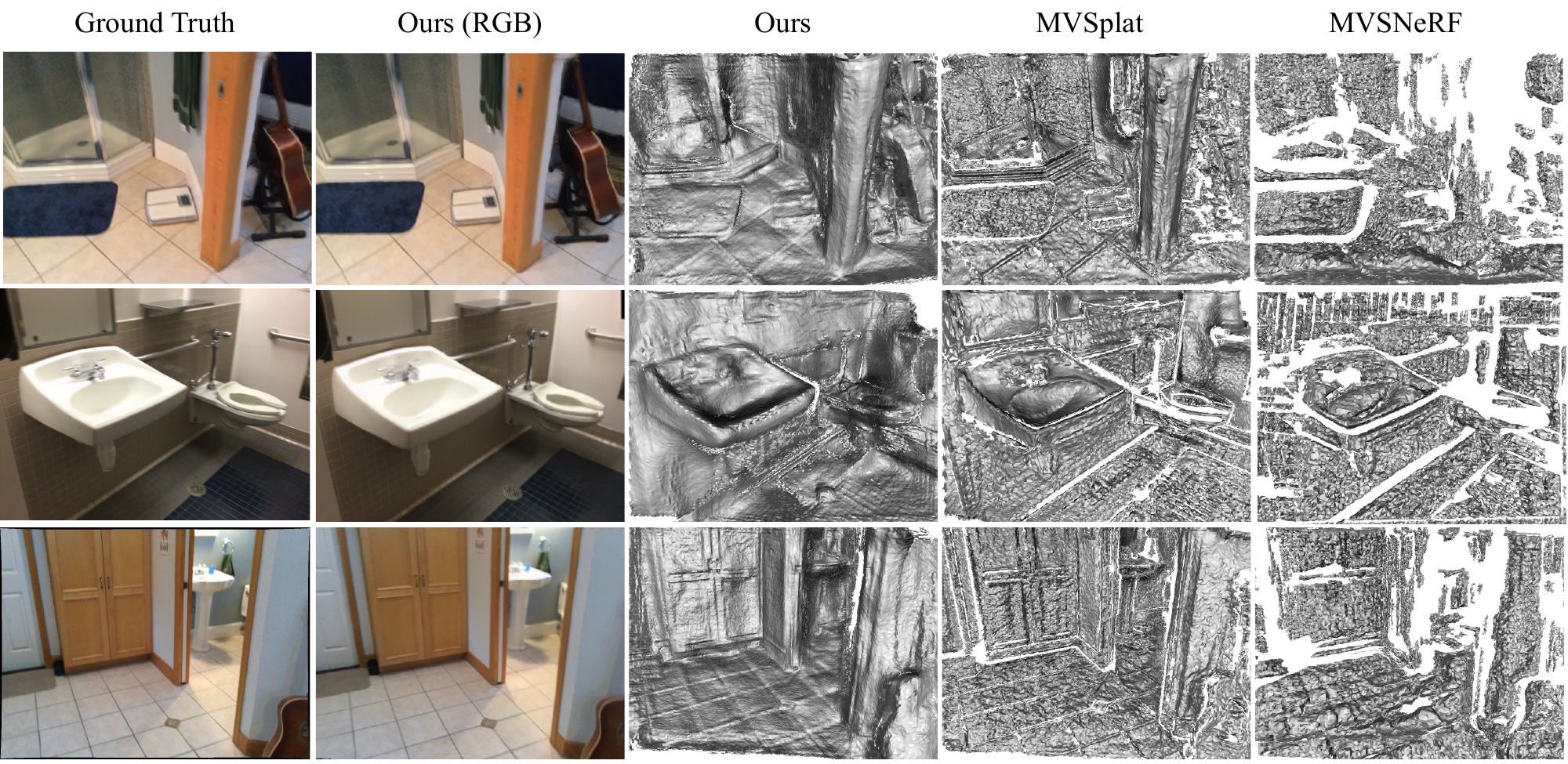}
  \caption{\textbf{Quanlitative Comparisons on Scannet Dataset.} While the baseline methods provide meshes with holes and uneven surfaces, MeshSplat successfully reconstruct the scene with smoother and more complete meshes.}
  \label{fig:main_scannet}
\end{figure*}

\section{Experiments}
\label{sec:exper}

\subsection{Settings}

\noindent\textbf{Datasets.} We train and evaluate our model on Re10K~\cite{zhou2018stereo} and Scannet~\cite{dai2017scannet} datasets. Re10K is a large-scale home walkthrough multi-view dataset obtained from YouTube, containing 67,477 training scenes and 7,289 testing scenes. Scannet is a multi-view real-world indoor dataset, and following \cite{zhang2022nerfusion}, we use 100 scenes for training and 8 scenes for testing. To evaluate the model generalizability, we also perform cross-dataset evaluations from Re10K to Scannet and Replica~\cite{straub2019replica}, a multi-view synthetic indoor dataset. Following \cite{zhi2021place}, we use 8 scenes for testing in Replica dataset.

\noindent\textbf{Implementation Details.} 
We employ different training strategies for Re10K and Scannet.
For experiments on Re10K, we crop the input images to $256\times256$ and train for 200,000 steps with a batch size of 12. For experiments on Scannet, we crop the input images to $512\times384$ and train for 75,000 steps with a batch size of 4. 

\subsection{Main Results}

\noindent\textbf{Mesh Reconstruction.} 
To enable comparisons with the state-of-the-arts, we transfer four methods from generalizable novel view synthesis to mesh reconstruction as our comparison.
For fair comparison, we also apply TSDF fusion for them except for SparseNeuS as it can extract meshes directly. 
Table~\ref{tab:re10k} presents the experimental results on the two datasets. The experiments demonstrate that MeshSplat significantly outperforms these methods in all metrics, indicating that our approach has strong capabilities in surface reconstruction with sparse-view input. 

Figures~\ref{fig:main_re10k} and~\ref{fig:main_scannet} show the visualizations on Re10K and Scannet. 
For the compared methods, their extracted meshes suffer from holes and uneven surfaces. In contrast, MeshSplat can accurately reconstruct the meshes of complex scenes with smoother surfaces, demonstrating that our model can capture scene structure well even with sparse-view inputs. More comparisons can be found in Appendix.

\begin{figure*}
  \centering
  \includegraphics[width=\linewidth]{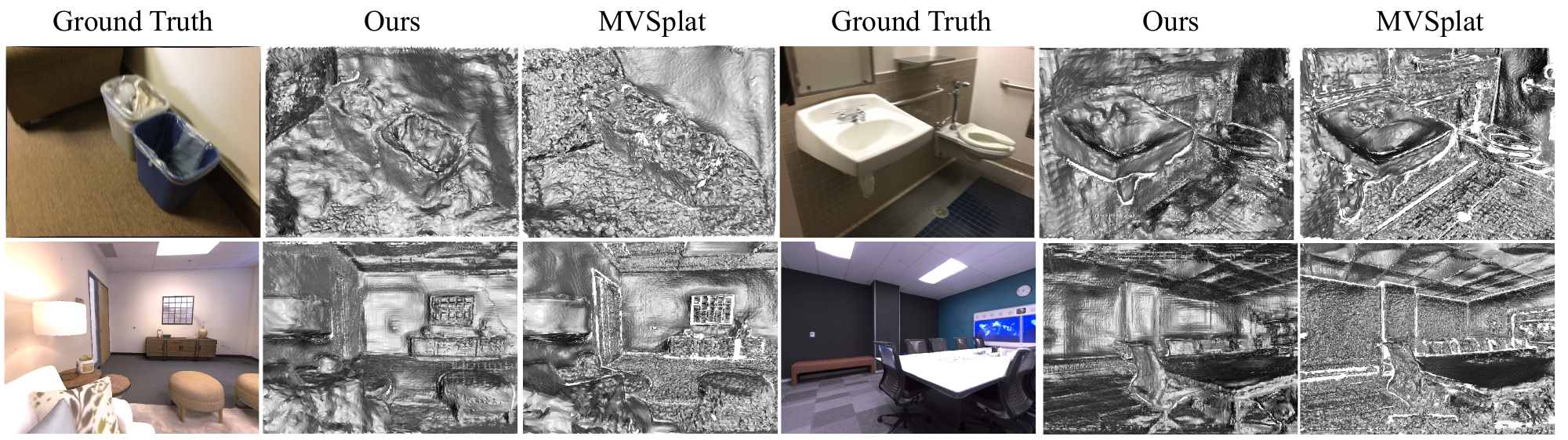}
  \caption{\textbf{Qualitative Comparisons in Zero-Shot Transfer Experiments on Scannet and Replica Datasets.} Compared to MVSplat, MeshSplat can still extract smoother surfaces, demostrating its generalization across different datasets.}
  \label{fig:zero_shots}
\end{figure*}

\noindent\textbf{Cross-Dataset Generalization.} 
To validate the generalization ability, We also conduct zero-shot transfer experiments on Scannet and Replica datasets respectively, using the model trained on the Re10K dataset. As shown in Table~\ref{tab:zero_shot} and Figure~\ref{fig:zero_shots}, our method still provides significant improvements compared to the baselines, demonstrating the satisfactory generalizability of MeshSplat across diverse scenarios. 

\begin{figure*}
  \centering
  \includegraphics[width=0.90\linewidth]{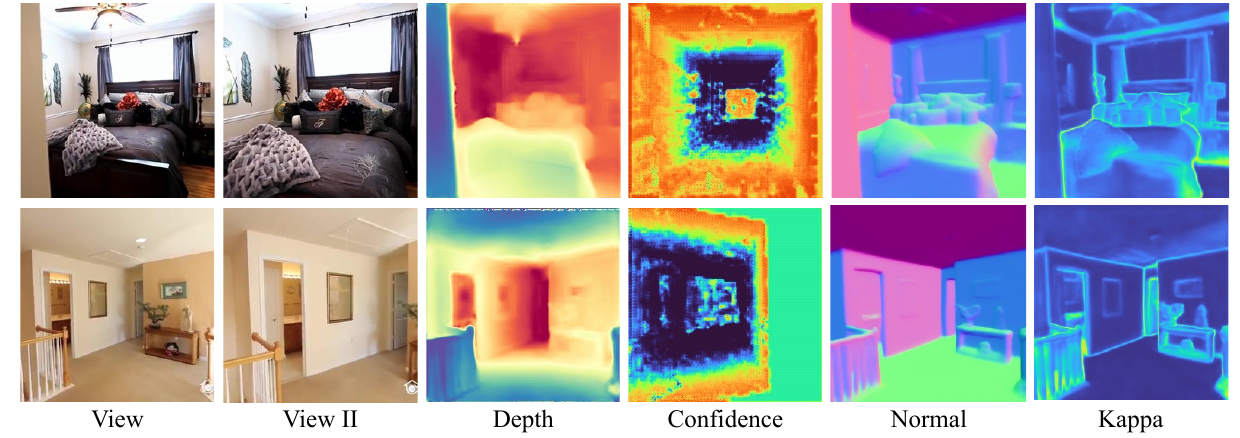}
  \caption{\textbf{Visualizations of output depth and normal maps, confidence maps used in WCD loss and kappa maps used in normal loss.} The confidence maps reflect the unconfident matching areas like texture-less areas and non-overlapped areas between the two views. For kappa maps, areas with higher uncertainty typically correspond to object boundaries.}
  \label{fig:main_geometry}
\end{figure*}

\noindent\textbf{Depth and Normal Maps Prediction.} 
To further show that MeshSplat can capture scene geometry accurately, we show the experiments on depth and normal maps prediction in Table~\ref{tab:depthnormal}. MeshSplat surpasses MVSplat by a large margin in all metrics, demonstrating that MeshSplat can better recover the scene geometry. We also show the visualizations of depth and normal maps in Figure~\ref{fig:main_geometry}, along with matching confidence maps in the WCD loss and kappa maps used in the normal loss. MeshSplat can accurately estimate depth and normal maps of the input views. The confidence maps clearly indicate regions of low confidence, such as texture-less areas and non-overlapped regions. For kappa maps, the higher uncertainty areas are typically object borders.

\subsection{Ablations}

\begin{table}[t]
\centering
\caption{\textbf{Results on Depth and Normal Prediction.} MeshSplat can predict depth and normal maps correctly.}
\begin{tabular}{llllll}
\toprule
\multicolumn{1}{c}{} & \multicolumn{2}{c}{Depth}                                                                      & & \multicolumn{2}{c}{Normal}                                                                   \\ \cmidrule{2-3} \cmidrule{5-6}
                      & \multicolumn{1}{c}{AbsRel↓} & \multicolumn{1}{c}{AbsDiff↓} & & \multicolumn{1}{c}{Mean↓} & \multicolumn{1}{c}{\textless{}30\textdegree↑} \\ \midrule

MVSplat               & 0.1692                    & 0.3197                   & & 57.16                & 0.1357                           \\ 
MeshSplat             & \textbf{0.0910}                   & \textbf{0.1680}                        & & \textbf{33.84}               & \textbf{0.6026}                           \\ \bottomrule
\end{tabular}

\label{tab:depthnormal}
\end{table}

\begin{table}[]
\centering
\caption{\textbf{Ablation Studies on Scannet Dataset.} NPN stands for the normal prediction network and NLL loss.}
\begin{tabular}{ccccc}
\toprule
 \# & 2DGS & \multicolumn{1}{c}{WCD Loss} & NPN & CD↓           \\ \midrule
  1 &   & \multicolumn{1}{c}{}         &                           & 0.3748 \\ 2 &
\checkmark    &                               &                           & 0.2948 \\ 3 &
\checkmark    & \checkmark                             &                           & 0.2769  \\ 4 &
\checkmark    &                               & \checkmark                         & 0.2642  \\ 5 &
\checkmark    & \checkmark                             & \checkmark                         & \textbf{0.2606} \\ \bottomrule
\end{tabular}

\label{tab:ablation}
\end{table}

\begin{figure}
  \centering
  \includegraphics[width=\linewidth]{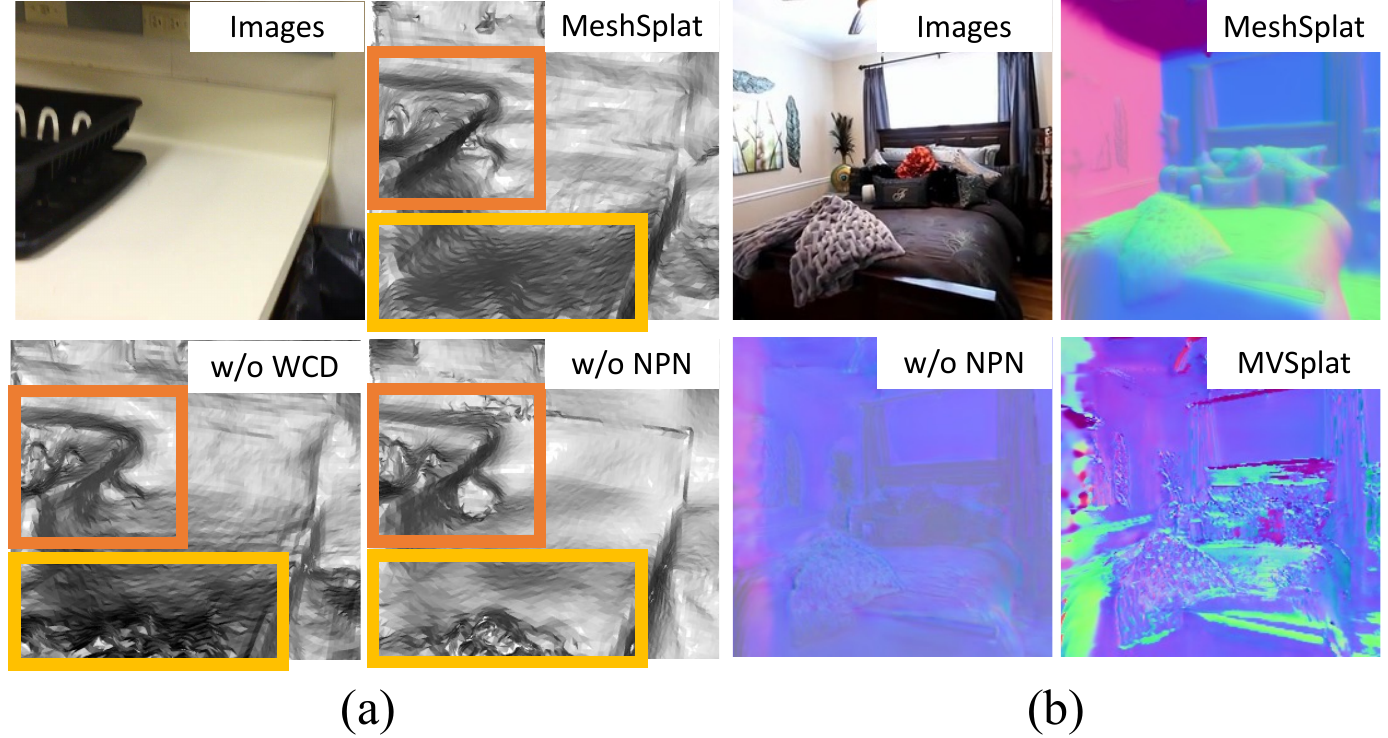}
  \caption{\textbf{Qualitative Ablation Studies.} (a) Ablation studies on proposed modules. (b) Comparisons of rendered normal maps by Gaussian Splatting.}
  \label{fig:ablation}
\end{figure}

\noindent\textbf{Main Components.} 
Table~\ref{tab:ablation} and Figure~\ref{fig:ablation} (a) present our ablation studies conducted on the Scannet dataset. The first two lines of the table indicate that 2DGS is more suitable for mesh reconstruction tasks compared to 3DGS. Moreover, by incorporating WCD loss and normal prediction network, 2DGS can align better with the actual surfaces. These enhancements significantly improve surface fidelity and reduce geometric artifacts in the final output. Overall, these results demonstrate that each proposed component plays a complementary role in boosting the reconstruction quality.

\noindent\textbf{Gaussian normals.} 
To further demonstrate the necessities of the normal prediction network, we present comparisons on normal maps rendered by Gaussian Splatting in Figure~\ref{fig:ablation} (b). Following~\cite{chen2023neusg}, we define the normal vector of 3DGS as the direction of the axis which has the minimum value of scales. 2DGS suffer from more significant degradation in normals compared to 3DGS when no regularization is applied. After incorporating the normal prediction network and its corresponding regularization loss, the predicted orientations of 2DGS become more accurate. 

\section{Limitations}
\label{sec:supp_limit}

Although our model achieves promising results in generalizable sparse-view surface reconstruction, it still has some limitations. MeshSplat sometimes may predict discontinuous depth maps in weakly textured areas, as shown in the Appendix, although the rendered RGB image is reliable. This is probably due to the ambiguity of feature matching in these areas. Additionally, MeshSplat cannot reconstruct surfaces in regions not visible from the input views. Incorporating generative approaches might be a good choice.

\section{Conclusion}
\label{sec:conclu}

We present MeshSplat, a generalizable surface reconstruction framework designed for sparse-view inputs. 
We use 2DGS as scene representation and construct an end-to-end surface reconstruction network. To better meet the needs for more accurate prediction of 2DGS, we propose the WCD loss to regularize coarse point maps, and also use a normal prediction network to predict the orientations of 2DGS. Experiments demonstrate that MeshSplat outperforms existing methods in generalizable sparse-view mesh extraction.

{
    \small
    \bibliographystyle{ieeenat_fullname}
    \bibliography{main}
}

\clearpage
\setcounter{page}{1}
\maketitlesupplementary

\section{Uncertainty-Guided Negative Log-Likelihood Loss in Normal Prediction}
\label{sec:supp_nll}

In this section, we explain the uncertainty-guided negative log-likelihood (UG-NLL) loss we use in the normal map supervision, which is proposed by \cite{bae2021estimating}. 

For each pixel $i$, with the predicted normal vector $n_i$, the actual normal vector $\hat n_i$ ideally follows the Angular von Mises-Fisher distribution:
\begin{equation}
  p_\mathrm {AngMF}(\hat n_i|n_i,\kappa_i)=\frac{(\kappa_i^2+1)\exp(-\kappa_i\cos^{-1}n_i^T\hat n_i)}{2\pi(1+\exp(-\kappa_i\pi))}
  \label{eq:angmf_prob}
\end{equation}
where $\kappa_i>0$ is the kappa value of this pixel. Higher $\kappa_i$ indicates a lower uncertainty in normal vector prediction. From this, we can compute the negative log-likelihood (NLL) loss:
\begin{equation}
\begin{split}
  \mathcal L_\mathrm {AngMF}(n_i,n_i^{\mathrm {gt}},\kappa_i)=-\log(\kappa_i^2+1)+\log(1+\\\exp(-\kappa_i\pi))+\kappa_i\cos^{-1}n_i^T n_{i}^{\mathrm {gt}}
\end{split}
\label{eq:supp_angmf_loss}
\end{equation}

In order to ensure efficient training, UG-NLL loss is only applied to several sampled pixels based on their $\kappa$. Specifically, we first sample the top $\beta\cdot N$ pixels with the lowest $\kappa$ at each scale. Then, we randomly sample $(1-\beta)\cdot N$ additional pixels, forming the set of all sampled pixels $P_\mathrm {sample}$.

\section{Datasets}
\label{sec:supp_datasets}

\begin{figure*}
  \centering
  \includegraphics[width=\linewidth]{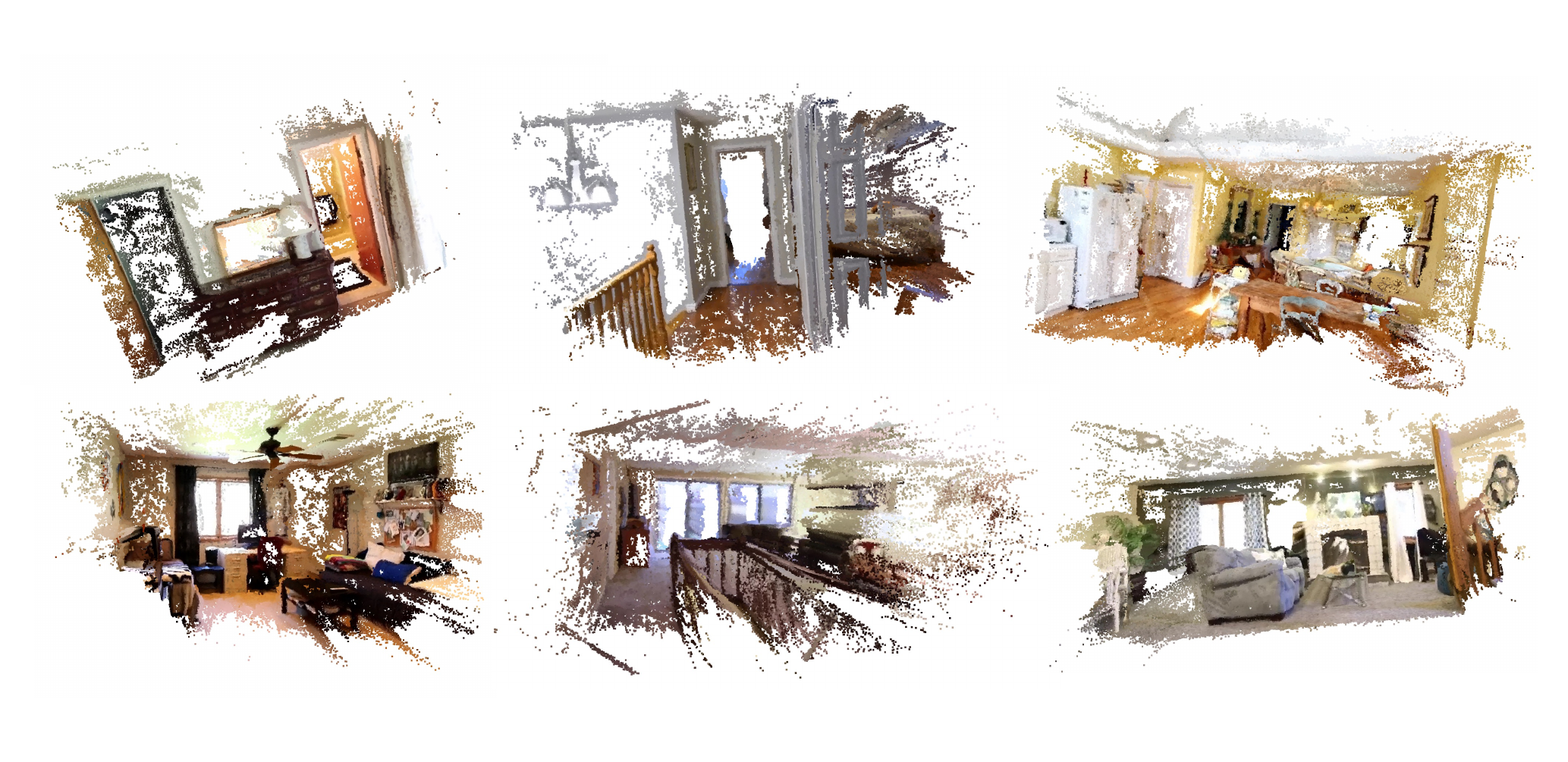}
  \caption{\textbf{Visualizations of our ground-truth dense point clouds provided by COLMAP in Re10K datasets.}}
  \label{fig:supp_colmap}
\end{figure*}

Our experiments are based on three indoor multi-view datasets, including Re10K~\cite{zhou2018stereo}, Scannet~\cite{dai2017scannet} and Replica~\cite{straub2019replica} datasets. RE10K is licensed by Google LLC under a Creative Commons Attribution 4.0 International License. Scannet is under MIT license. Replica is under a custom license that only allows a research or educational purpose. Since Re10K does not have ground-truth meshes or point clouds, we use COLMAP~\cite{schonberger2016colmap} to reconstruct dense point clouds for 20 scenes under dense-view inputs with provided camera parameters, and take these point clouds as ground truth. Additionally, since sparse input viewpoints cannot observe the entire content of the scene, we cull ground-truth meshes or point clouds based on the input view frustums. Visualizations of some of these point clouds can be found in Figure~\ref{fig:supp_colmap}.

\section{Additional Implementation Details}
\label{sec:supp_implement}

\noindent\textbf{Network Architecture.} The backbone of our model follows the same structure and initialization methods as MVSplat~\cite{chen2024mvsplat}. For the cost volume construction, we sample $D=128$ depth candidates between the near and far planes. For the normal prediction network, we use a total of 10 convolutional layers to predict the 2D Gaussian orientations, and the pseudo ground-truth supervision is applied to normal maps at all scales including 1/4, 1/2 and full resolution. For the depth refinement network, they are implemented as U-Net with self-attention and cross-attention layers.

\noindent\textbf{Training Strategy.} We use the Adam optimizer with a maximum learning rate of $2\times 10^{-4}$ to train our network. The weights of all components of the loss function are set as follows: $w_1=1.0$, $w_2=5.0\times 10^{-3}$, $w_3=5.0\times 10^{-3}$, $w_{11}=1.0$, $w_{12}=0.1$ and $\alpha=0.1$. For the uncertainty-guided random sampling applied in normal prediction network, we set $\beta=0.7$, $N=0.4\times H\times W$, where $(H, W)$ is the image size at the current processing scale. All experiments are performed on a single NVIDIA A800 GPU.

\noindent\textbf{Others.} On the Re10K dataset, we set the camera near and far planes to $d_{near}=1.0$ and $d_{max}=100.0$, and we set the voxel size to 0.005 and the truncated threshold to 0.1 for TSDF fusion. For the Scannet and Replica datasets, we set the camera near and far planes to $d_{near}=0.5$ and $d_{max}=15.0$, and the voxel size and truncated threshold to 0.01 and 0.08 for TSDF fusion.

\section{Additional Experimental results}
\label{sec:supp_exper}

\begin{table}[]
\centering
\begin{tabular}{@{}ccc@{}}
\toprule
\multicolumn{1}{c}{} & Time (s) & Params (M) \\ \midrule
MVSNeRF*~\cite{chen2021mvsnerf}              & 0.76     & 0.341   \\
SparseNeuS*~\cite{long2022sparseneus}           & 7.0483   & 0.843      \\
pixelSplat~\cite{charatan2024pixelsplat}           & 0.194    & 125.4      \\
MVSplat~\cite{chen2024mvsplat}              & 0.072    & 12.0         \\
MeshSplat            & 0.102    & 13.3       \\ \bottomrule
\end{tabular}
\caption{\textbf{Model Efficiency Analysis. } We report rendering time and model sizes of MeshSplat and baseline methods.  * denotes that these methods render only one image per forward pass.} 
\label{tab:supp_effi}
\end{table}

\noindent\textbf{Model Efficiency Analysis. } We report the rendering time and the size of the models in Table~\ref{tab:supp_effi}. All experiments are conducted on a single NVIDIA 3090 GPU. Compared with NeRF-based methods, MeshSplat achieves significantly faster rendering speeds. Additionally, MeshSplat has only a few extra model parameters and little extra rendering time compared to MVSplat.

\begin{figure*}
  \centering
  \includegraphics[width=0.95\linewidth]{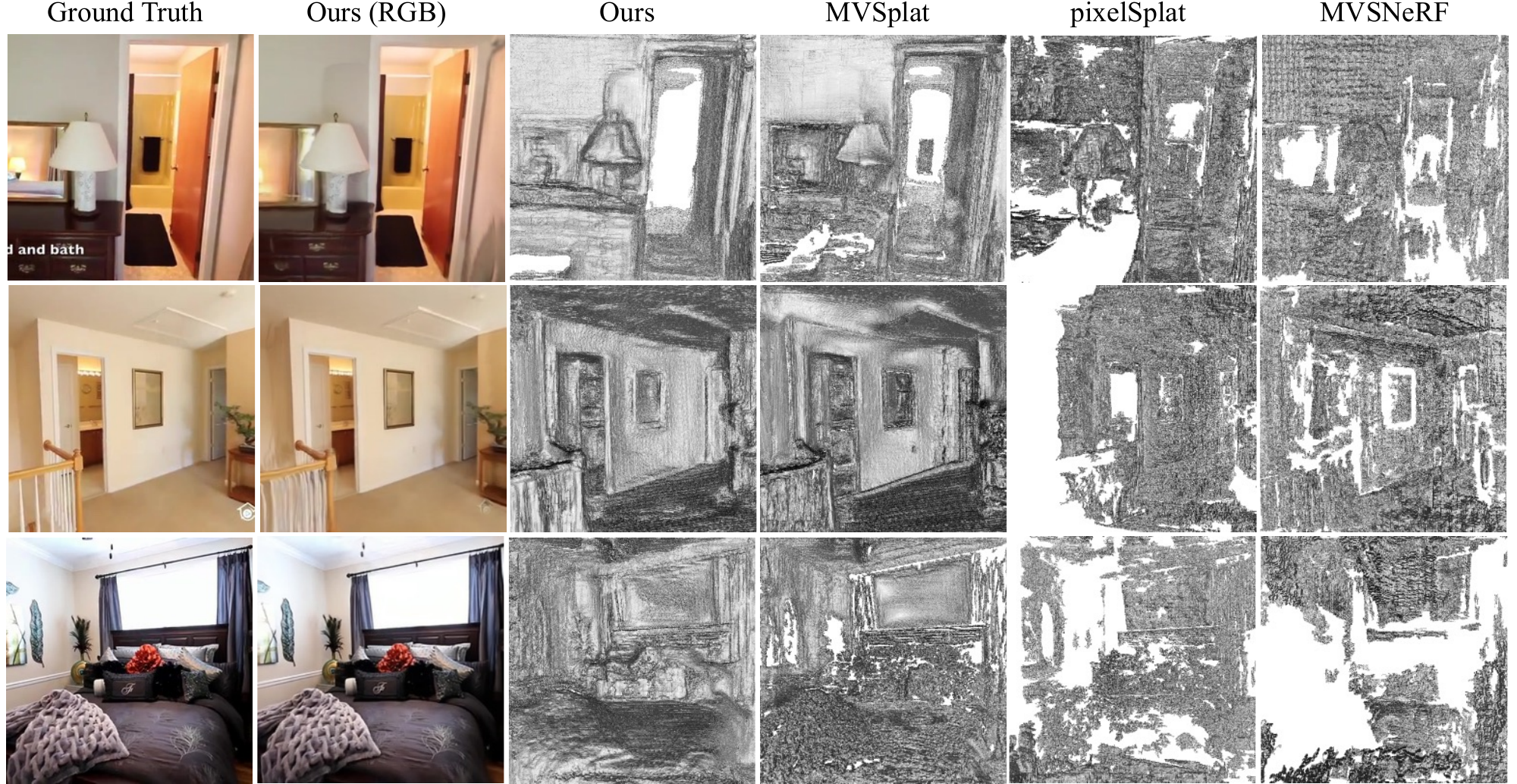}
  \caption{\textbf{Additional Visualizations of Extracted Meshes in Re10K Datasets.}}
  \label{fig:supp_re10k}
\end{figure*}

\begin{figure*}
  \centering
  \includegraphics[width=0.95\linewidth]{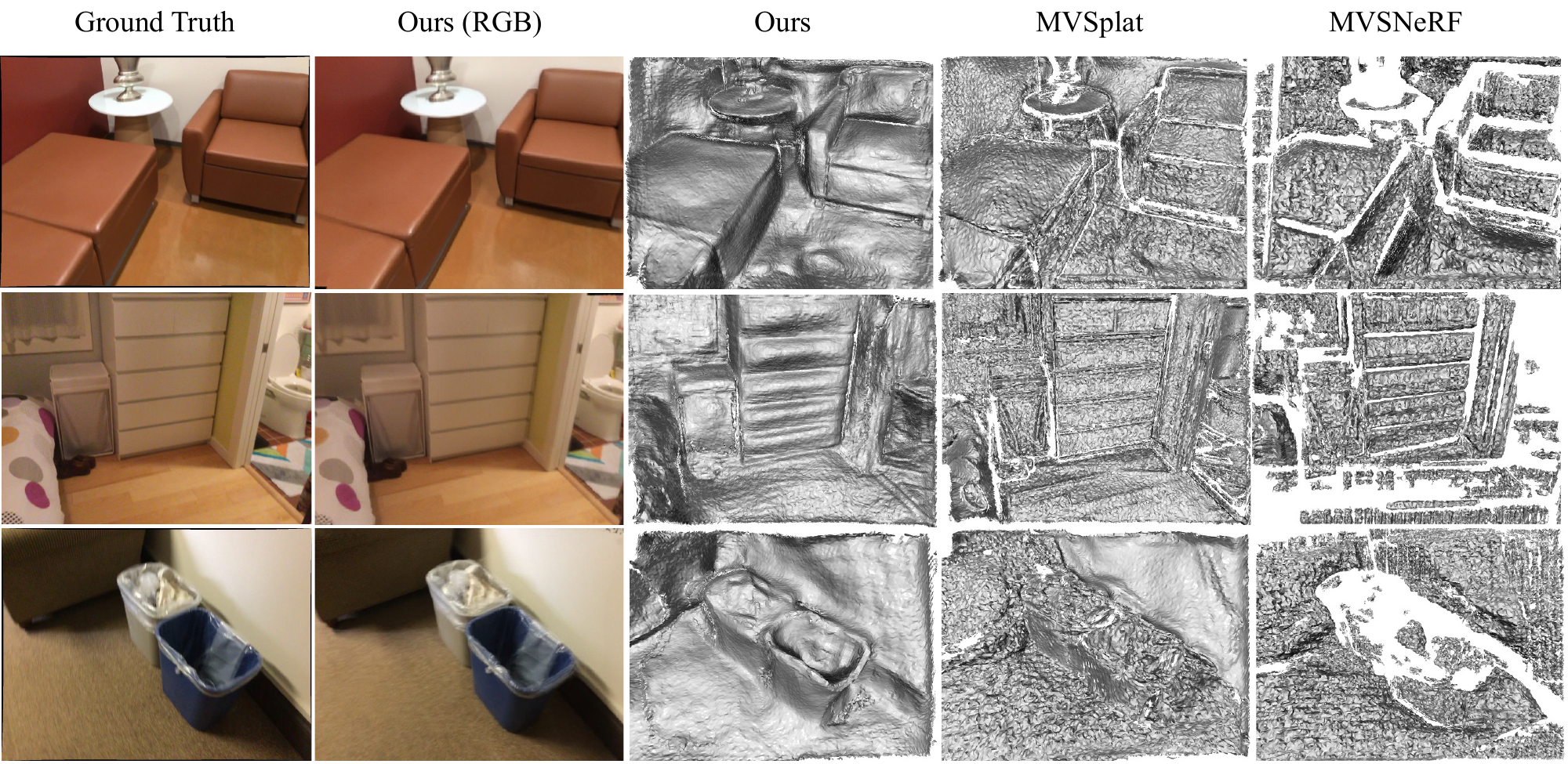}
  \caption{\textbf{Additional Visualizations of Extracted Meshes in Scannet Datasets.}}
  \label{fig:supp_scannet}
\end{figure*}

\begin{figure*}
  \centering
  \includegraphics[width=0.95\linewidth]{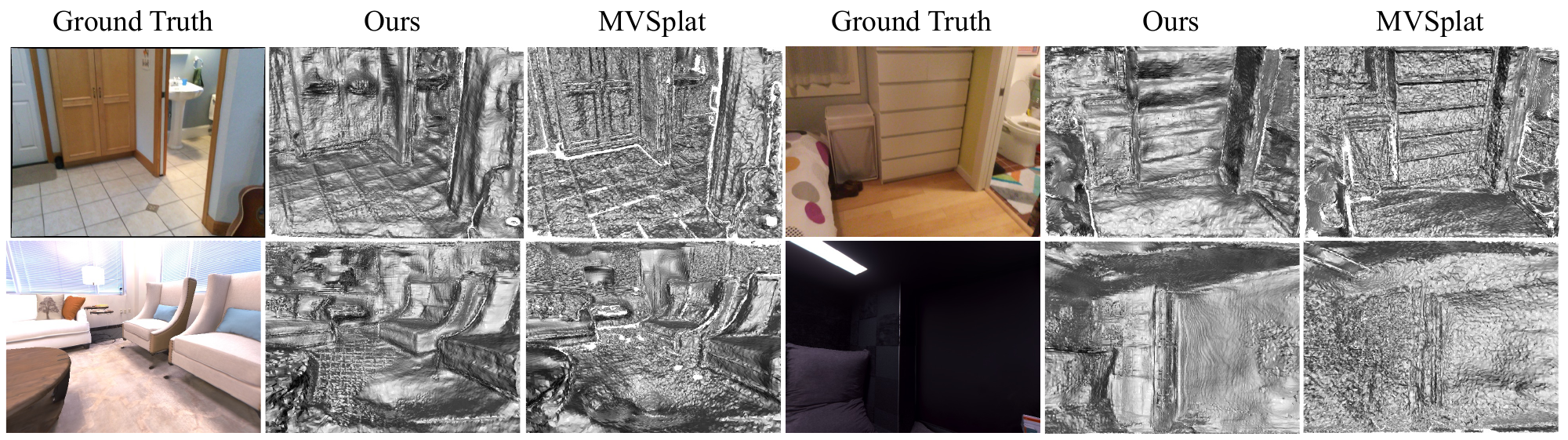}
  \caption{\textbf{Additional Visualizations of Extracted Meshes in Cross-Dataset Generalization Experiments.}}
  \label{fig:supp_zero_shot}
\end{figure*}

\noindent\textbf{Additional Mesh Visualization. } Figure~\ref{fig:supp_re10k}, \ref{fig:supp_scannet} and \ref{fig:supp_zero_shot} show additional visualizations of reconstructed mesh. Please refer to corresponding captions of these figures for details.

\begin{figure*}
  \centering
  \includegraphics[width=0.8\linewidth]{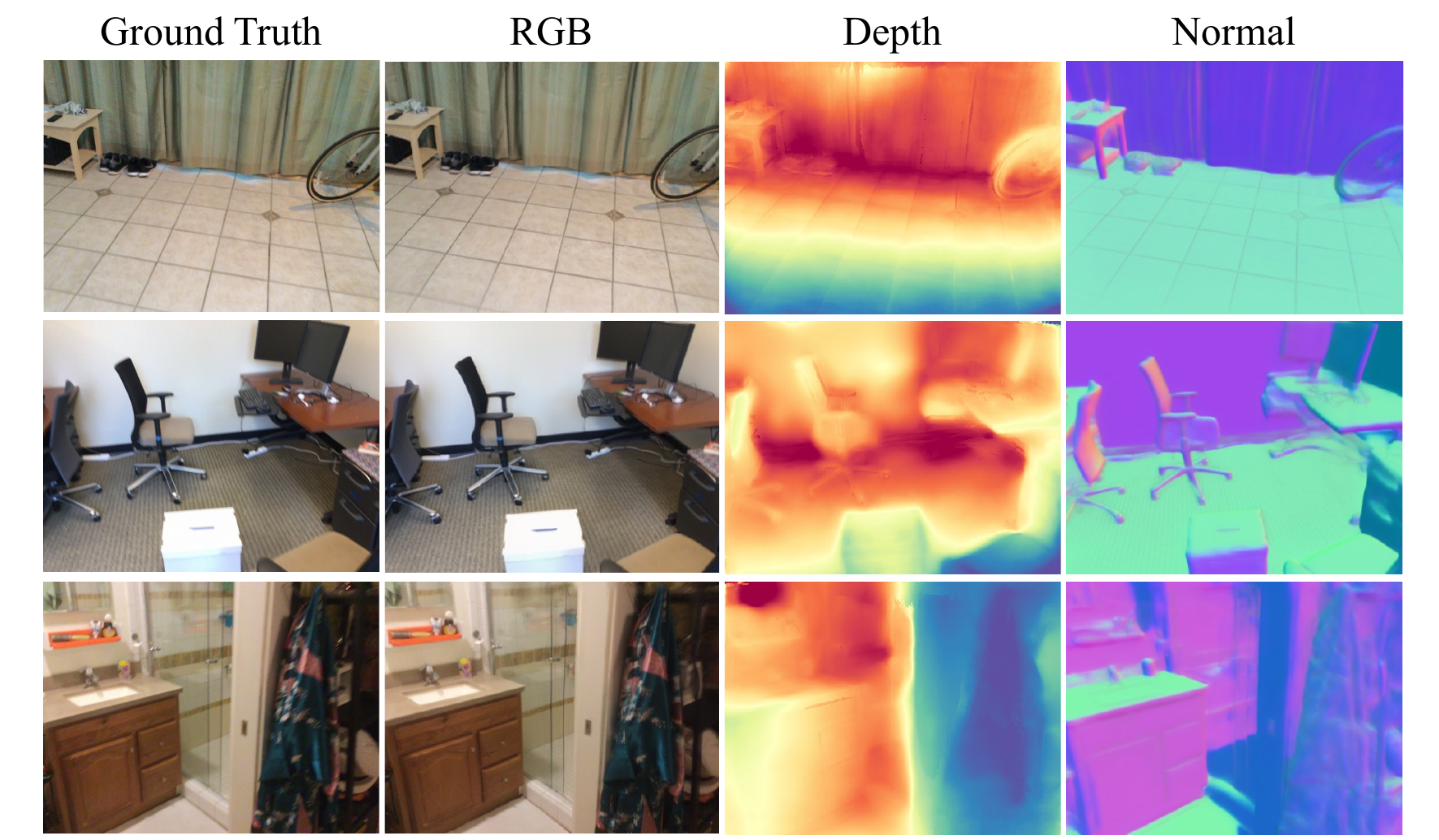}
  \caption{\textbf{Visualizations of Rendered RGB Images, Depth Maps and Normal Maps in the Scannet Dataset.}}
  \label{fig:supp_geometry}
\end{figure*}

\begin{figure*}
  \centering
  \includegraphics[width=0.8\linewidth]{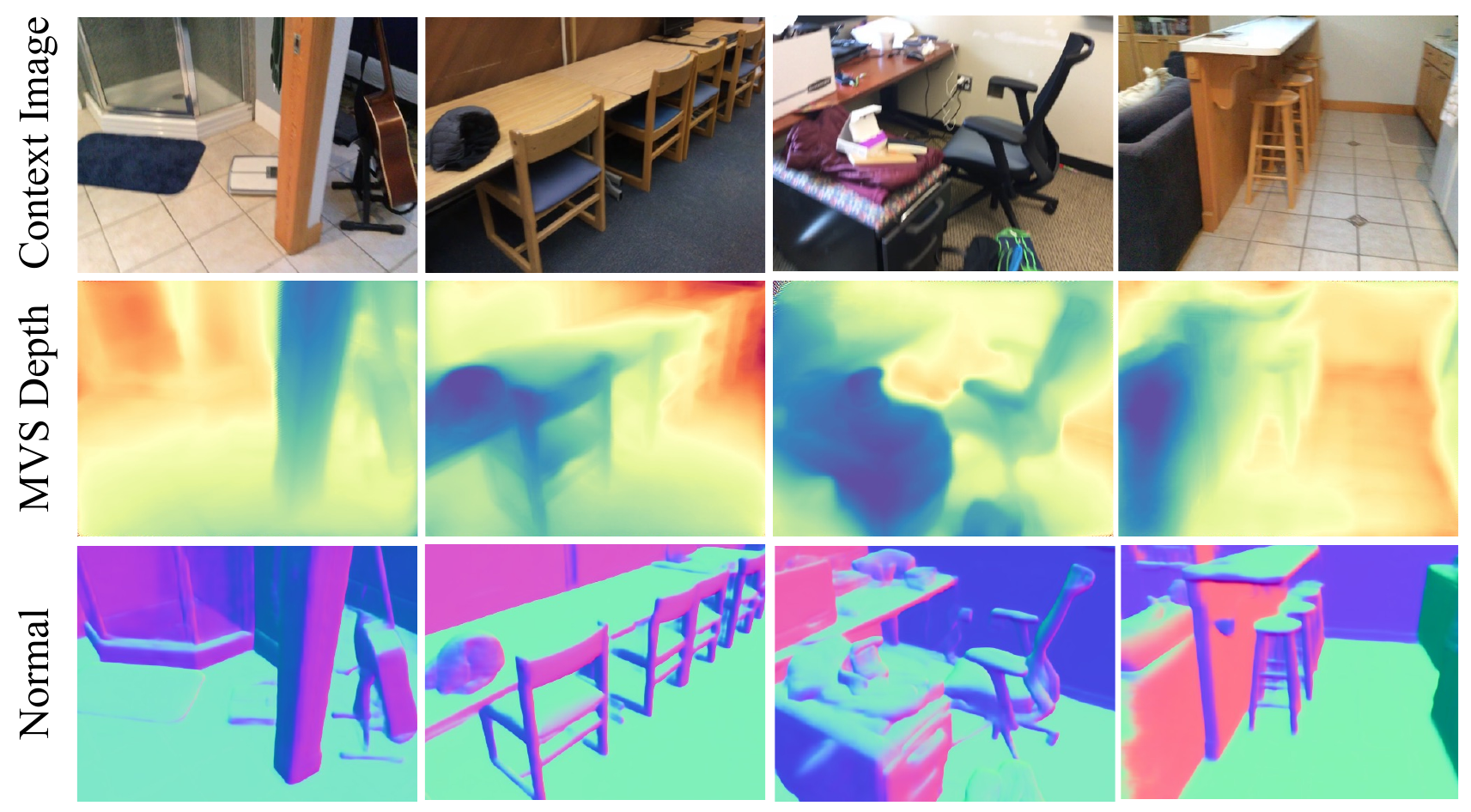}
  \caption{\textbf{Predicted Depth Maps and Normal Maps of the Gaussian Prediction Network.}}
  \label{fig:supp_kappa}
\end{figure*}

\noindent\textbf{Additional Depth and Normal Visualization. } Figure~\ref{fig:supp_geometry} shows the rendered depth maps and normal maps by 2DGS in the Scannet dataset. Additionally, we show depth maps and normal maps of context images predicted by the Gaussian prediction network in Figure~\ref{fig:supp_kappa}.

\section{Video Demo}
\label{sec:supp_video}

We apply a video demo of MeshSplat to further show the mesh reconstruction results. Please refer to the supplementary materials for details.

\begin{figure*}
  \centering
  \includegraphics[width=0.8\linewidth]{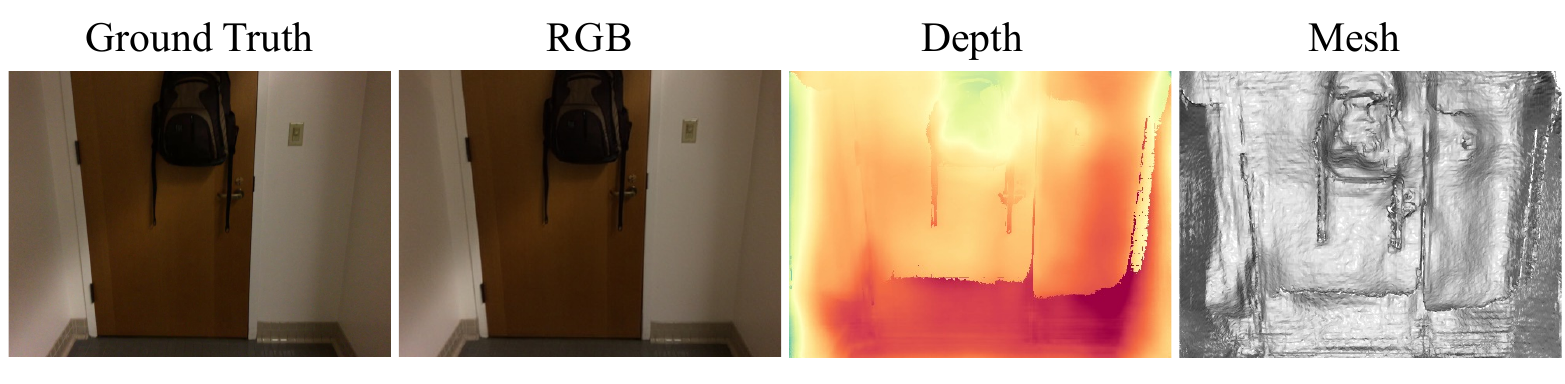}
  \caption{\textbf{Failure Cases of MeshSplat.}}
  \label{fig:supp_fail}
\end{figure*}

\end{document}